\documentclass[11pt]{article}

\usepackage[T1]{fontenc}
\usepackage[utf8]{inputenc}
\usepackage{lmodern}
\usepackage[margin=1in]{geometry}
\usepackage{microtype}
\usepackage{graphicx}
\usepackage{booktabs}
\usepackage{longtable}
\usepackage{array}
\usepackage[numbers,sort&compress]{natbib}
\usepackage[hidelinks]{hyperref}

\makeatletter
\def\bstctlcite{\@ifnextchar[{\@bstctlcite}{\@bstctlcite[@auxout]}}
\def\@bstctlcite[#1]#2{\@bsphack
  \@for\@citeb:=#2\do{%
    \edef\@citeb{\expandafter\@firstofone\@citeb}%
    \if@filesw\immediate\write\csname #1\endcsname{\string\citation{\@citeb}}\fi}%
  \@esphack}
\makeatother

\newcolumntype{L}[1]{>{\raggedright\arraybackslash}p{#1}}
\newcommand{\code}[1]{\texttt{#1}}
\newcommand{\PaperTitle}{EA-Graph: Artifact-Anchored Verification Memory for Coding Agents under Upstream Drift}
\newcommand{\PaperAuthors}{%
  Hwai-Jung Hsu\textsuperscript{1}\href{https://orcid.org/0000-0002-2586-4484}{\textsuperscript{\scriptsize ORCID}},
  Cheng-Jan Chi\textsuperscript{2}, and
  Hanna Everett\textsuperscript{2}\\[0.55em]
  \small\textsuperscript{1}Sustainable and Peer Education Center,
  National Yang Ming Chiao Tung University\\
  \small\textsuperscript{2}Office of AI Affairs,
  National Yang Ming Chiao Tung University\\
  \small Hsinchu, Taiwan, R.O.C.\\
  \small\href{mailto:hjhsu@nycu.edu.tw}{hjhsu@nycu.edu.tw}%
}
\newcommand{\PaperDate}{August 2026}

\title{\PaperTitle}
\author{\PaperAuthors}
\date{\PaperDate}

\begin{document}
\bstctlcite{eagraph:BSTcontrol}
\maketitle

\begin{abstract}

Coding agents increasingly perform software work across multiple sessions. Prose notes can record that a claim was checked without preserving the program state used in that check. After an upstream change, a project may still build even when an earlier verification claim is no longer valid.

We present \textbf{EA-Graph}, an artifact-anchored memory for verification claims. It represents artifacts as first-class nodes at sub-path granularity and resolves aliases to their leaf definitions. Each claim is anchored to the content used to establish it. Evidence strength remains separate from freshness, so a claim over changed support can be withdrawn. If the required replacement content is unavailable, EA-Graph records the claim as unprovable rather than forcing a guess.

The evaluation uses generated repositories with behavior-to-artifact ground truth known by construction. Each world contains 96 behaviors across 12 modules, a readable but non-executable reference, and an unversioned upstream drop. The drop includes value drift, logic drift, and deliberately withheld content. The task is to classify prior claims as unaffected, affected, or unprovable.

The analysis covers \textbf{42 sessions} across seven clean worlds, 14 model-world instances, three memory conditions, and two model tiers. In the Haiku round, the anchored condition outscored both prose notes and no persistent memory in all seven worlds. Each exact paired Wilcoxon comparison yielded \emph{p} = 0.0156. In the Sonnet round, the anchored condition was perfect, but frequent control ceilings left the preregistered contrasts non-significant. No session fabricated withheld content. The results support a bounded claim: artifact-anchored memory improved the smaller model's provability judgment in this testbed. An exploratory cross-model pattern further suggests that structured claim memory may narrow a capability gap by externalizing in-session re-derivation. This pattern motivates a hypothesis; it does not establish cross-model equivalence. The study makes no claim about efficiency or repair quality.

\end{abstract}

\newpage

\section{Introduction}\label{sec:1}

Large language model agents increasingly perform software engineering tasks such as repository exploration, implementation, testing, and repair. Realistic development, however, often spans multiple sessions or agents. A later agent must continue from decisions, observations, and verification results produced earlier, commonly through progress summaries or prose notes. These records describe development history but do not preserve enough information to decide whether prior work remains valid after the software changes.

This limitation becomes consequential when software evolves between handoffs. A note may describe what was implemented and why, yet omit the precise code, data, configuration, or external assumptions on which the work depends. Prose notes record \emph{where} a claim was checked; they do not record \emph{what it was checked against}. The distinction remains invisible until the upstream changes.

Consider a concrete case. A session verifies that a ported behavior matches its reference and records the result. Later, a revised reference arrives. No file is added or removed, every import still resolves, and the project compiles. Yet one entry in a lookup table now holds a different value. The verified behavior reads that entry, so the recorded claim is now false. Nothing in the workspace reveals the change in validity.

\textbf{What is missing announces itself; what is stale does not.} A deleted file, broken import, or renamed symbol produces an error that can stop the build. A changed value can instead produce a silent error in any behavior that reads it. A file-level view discards the information needed to identify those behaviors. In the worlds studied here, file-granularity analysis marks about 88 of 96 behaviors as suspect when only about 17 are affected.

\begin{figure}[t]

\centering

\includegraphics[width=0.62\textwidth]{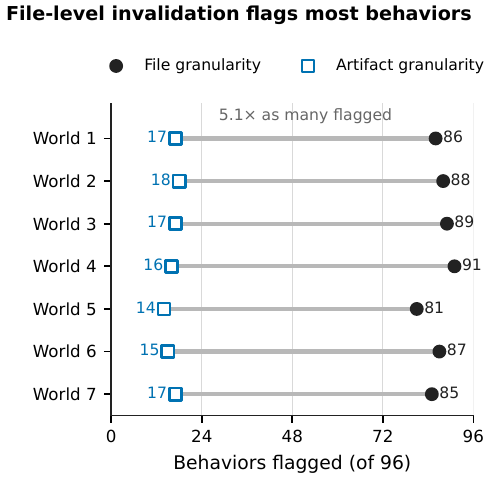}

\caption{Artifact- and file-level invalidation across seven worlds.}

\label{fig:figure1_over_invalidation}

\end{figure}

Figure~\ref{fig:figure1_over_invalidation} quantifies this mismatch across the seven clean worlds. The artifact-level bars show behaviors whose claims require review. The larger file-level bars include false alarms caused by collapsing distinct table entries and other sub-path artifacts into their containing files. This gap motivates EA-Graph's use of sub-path identity.

\subsection{Motivating practitioner case}

The work was motivated by a practitioner's longitudinal migration of a character asset between game-engine generations. The migration spanned many agent-driven sessions. Two rounds of its most difficult material-migration stage produced 16 outputs without an accepted result. The maintainer then constructed a \textbf{project-specific artifact evidence index} that linked code, assets, sources, known gaps, and machine-readable checks. Its accepted snapshot contained 6,311 entities, 16,994 relations, and five explicit gaps. The records distinguished direct support, partial support, unresolved relationships, and boundaries at which engine-native behavior could not be inspected.

A later audit detected two source-lock mismatches and marked the index \code{STALE}. Previously accepted artifacts remained usable, but the affected evidence required review or reconstruction. It could no longer be treated as current. This case motivates durable artifact identity, evidence boundaries, and freshness tracking. Because it was not a controlled study, it supports no causal, efficiency, or performance claim.

\subsection{Problem statement}

The operative question at the start of a session, after an upstream has moved, is:

\begin{quote}

\textbf{Which of the claims I previously verified can I no longer stand behind, and which of those can I not settle at all?}

\end{quote}

This question requires the granularity of the values and logic that a behavior uses; file granularity is insufficient. It also has three possible outcomes. A claim may be unaffected, affected, or \textbf{unprovable}. The third outcome applies when a dependency changed but its new content was not delivered. No evidence available to the agent can then settle the claim. A system with only the first two outcomes forces a guess precisely where guessing is most harmful.

\subsection{Method overview}

EA-Graph stores verification claims over first-class artifacts with sub-path identity. Each claim is anchored to the specific artifact content used in its verification. Evidence strength is tracked separately from freshness, so a claim over drifted support is withdrawn rather than merely qualified (Section~3).

The evaluation uses generated repositories rather than an existing open-source project. Existing projects could introduce training-data exposure, incomplete ground truth, and visible version-control history. Each world contains 96 behaviors across 12 modules and a reference that the agent can read but not execute. The upstream arrives as an unversioned drop with some content deliberately withheld (Section~4). An accompanying shortcut catalog documents the routes that had to be closed. Otherwise, the evaluation could measure access to leaked answers rather than agent memory.

The evaluation comprises 42 analyzed sessions across seven clean worlds and 14 model-world instances. Each session uses one of three memory conditions: EA-Graph claims anchored to artifact content (\textbf{ANCHOR}), prose verification notes (\textbf{PROSE}), or no persistent record (\textbf{NONE}). In the smaller-model round, ANCHOR outscored both controls in all seven worlds. In the larger-model round, ANCHOR was perfect in all seven worlds, but frequent ceiling scores in the controls left the preregistered comparisons inconclusive. Across the shared worlds, the smaller model with ANCHOR nevertheless approached the larger model's descriptive performance. The smaller model's controls did not. This post-hoc, unpaired comparison is not an equivalence test. It motivates a research hypothesis rather than establishing an effect. Structured verification-claim memory may narrow the observed capability gap by externalizing artifact re-derivation that the larger model performed within the session (Section~5--Section~6).

\subsection{Contributions}

\begin{enumerate}

\item \textbf{EA-Graph} (Section~3): an artifact-centric verification memory with sub-path identity, alias resolution to leaf definitions, content anchors, independent representations of evidence and freshness, and an explicit unprovable state when changed support is unavailable.

\item \textbf{A generated testbed} (Section~4) in which the correct answer is known by construction and unreachable by shortcut, together with the catalog of shortcuts that had to be closed to make that true.

\item \textbf{An empirical result with a testable extension} (Section~5): anchored verification memory improves the smaller model's provability judgment over prose notes and no persistent memory in this testbed. An exploratory cross-model pattern further suggests that structured verification-claim memory may narrow the observed gap to a more capable model. A future paired equivalence design should test this possibility.

\item \textbf{Measurement lessons} (Section~6): file-granularity invalidation is quantified against artifact-granularity truth. The results also show that a methodology embedded only in a tool is ineffective when the task is posed at the wrong granularity. Finally, net repair scores systematically penalize the refusal discipline advocated here.

\end{enumerate}

The rest of the paper is organized as follows. Section~2 places the work among existing agent memories, repository representations, and program analyses. Section~3 defines EA-Graph and marks which of its parts this paper evaluates. Section~4 argues for the generated testbed and describes it. Section~5 reports the experiments. Section~6 discusses what the results mean, where the measures were wrong, and the threats to validity. Section~7 concludes.

\section{Related Work}\label{sec:2}

\subsection{Memory and externalization for agents}\label{sec:2.1}

Agent memory commonly uses a hierarchy of stores with policies for retention, summarization, and retrieval. MemGPT is a canonical example \citep{memgpt2023}. Recent surveys map the mechanisms and evaluations in this area \citep{memsurvey2026}. They also place memory alongside skills, protocols, and harness design as forms of externalization \citep{externalization2026}. This literature primarily addresses \emph{capacity}: what to retain when not everything fits in context. The present concern is orthogonal and logically prior. A retained item can be complete, well summarized, and correctly retrieved, yet become false because its subject has changed. Existing designs do not bind a retained claim to the content used to verify it and therefore cannot detect this failure.

The closest work in this line grades retained information through a provenance-aware, tiered memory. It links verified findings to their raw sources \citep{tiermem2026}. This design shares the motivation behind EA-Graph's evidence grades but applies it to general agent memory rather than code artifacts. EA-Graph instead retains claims about \emph{behaviors} bound to sub-path artifact identities. It also treats freshness as an independent axis with a refusal rule: a stale claim is withdrawn rather than returned for the agent to weigh.

\subsection{Repository representations and program analysis}\label{sec:2.2}

A parallel line gives agents structured repository views. CodexGraph exposes a queryable code graph \citep{codexgraph2025}, while RepoGraph uses a repository graph to guide repair \citep{repograph2025}. CodePlan frames repository-wide change as planning over an incrementally maintained dependency graph \citep{codeplan2024}. Practical tools also build tree-sitter knowledge graphs for exploration \citep{codebasememory2026}. These schemas represent \emph{code symbols}, including definitions, references, and calls. A lookup table is a symbol, but an individual entry read by a behavior is not. Value drift can occur at precisely that finer granularity. These schemas also lack evidence grades and freshness states. They cannot express or withdraw the claim that specific content was verified.

Program analysis can supply finer precision. Agent-facing systems now expose intra-procedural slicing to repair agents \citep{arise2026} and use path-wise data-flow facts for repository auditing \citep{repoaudit2025}. As Section~3.1 explains, this work occupies a complementary layer. Value-flow analysis answers questions within a procedure's value graph. The effect relation instead tracks dependencies across persistent system state. The two layers can compose. The present claim is limited to the second layer and its identity and evidence semantics.

Incremental analysis under evolving specifications predates agentic software engineering. Hsu and Wang update resource-conflict analysis as workflow specifications change \citep{hsu2008incremental}. EA-Graph shares the concern with avoiding full re-analysis after change, but operates on a different object: the validity of verification claims anchored to program artifacts, rather than resource conflicts in a formal workflow model. It provides intellectual lineage, not prior evidence for the present method.

\subsection{Multi-agent conflict and benchmark integrity}\label{sec:2.3}

A third line addresses coordination among concurrent agents. Examples include admission control before writes \citep{atm2026}, process-aware conflict detection \citep{scf2026}, and a large-scale study of merge conflicts in agent-authored pull requests \citep{agenticflict2026}. This work defines an important boundary. Those systems arbitrate between agents acting concurrently. The present failure instead involves one agent acting across time after an upstream changes between sessions. This study does not examine concurrency.

Finally, the release strategy of Section~4.5 responds to benchmark contamination. LiveCodeBench reduces this risk by continuously collecting newly published programming problems \citep{livecodebench2024}. The present setting poses a related release problem. Publishing generated repositories with their hidden answers can expose later evaluations to retrieval or memorization. The release therefore separates public design and protocol materials from request-only world contents and answers, as described in Section~4.5.

\section{EA-Graph}\label{sec:3}

EA-Graph stores verification claims about a codebase and the evidence supporting each claim. The question in Section~1 asks which earlier claims no longer hold and which cannot be settled. Answering it requires three properties absent from ordinary code representations. These properties define the model.

The complete state is written as:

\begin{quote}

\textbf{M = (G, C, ANCH, META, DISP)}, with \textbf{G = (V, A, OPS)}.

\end{quote}

Here \textbf{G} is the code--artifact effect graph defined in Section~3.1, and \textbf{C} is the set of verification claims. \textbf{ANCH} binds each claim to a canonical artifact identity and a digest of the content used in its verification. \textbf{META} assigns an independent \code{(evidence, freshness)} pair to stored facts and anchors. \textbf{DISP} records whether the maintained artifact should be retained or withdrawn. A withdrawal query checks ANCH against the current artifacts and returns \textbf{unaffected}, \textbf{affected}, or \textbf{unprovable}. DISP remains separate from this three-way outcome. Loss of proof does not authorize destruction of the last verified artifact.

A claim must attach to \textbf{the specific values and logic a behavior uses}, not to their containing files (Section~3.1). The attachment must \textbf{detect changes in the underlying content} (Section~3.2). The outcome must also permit \textbf{"this cannot be established"} rather than force a guess (Section~3.3). Section 3.4 states the queries served by the model. Section 3.5 describes the reduced implementation and the elements evaluated in this study.

\subsection{Artifacts as first-class nodes}\label{sec:3.1}

The graph component of an EA-Graph is a pair of node sets with a labelled effect relation between them:

\begin{quote}

\textbf{G = (V, A, OPS)}, where \textbf{V} are code nodes, \textbf{A} are artifact nodes, V \ensuremath{\cap} A = \ensuremath{\emptyset}, and \textbf{OPS \ensuremath{\subseteq} V \ensuremath{\times} A \ensuremath{\times} \{read, write, kill\}}.

\end{quote}

\textbf{Code nodes (V)} are function-level units. This granularity is a deliberate floor rather than a limitation of the formalism. Agent edits are often function-sized, and questions such as \emph{which behavior depends on this table?} do not require basic-block resolution.

\textbf{Artifact nodes (A)} are the persistent things code operates on: a configuration key, a database column, a lookup table, an asset GUID, a re-exported constant. Their identity is a canonicalized triple

\begin{quote}

\textbf{(store, path, subpath)}

\end{quote}

produced by a \emph{single} normalization function. \code{store} identifies the type of persistent medium, such as a Python module attribute, JSON document, or database schema. \code{path} locates the container, and \code{subpath} locates an item \emph{within} the container. Sub-path identity distinguishes this model from file-level representations. Two constants in the same module are separate artifacts, so a change to one does not affect the other. A file-level model collapses them into one unit and marks every reader of that file as suspect. Section 5 measures the resulting over-invalidation.

Identity carries a \textbf{subsumption} order. A whole document subsumes its keys, a table subsumes its entries, and sibling keys are disjoint. Queries can therefore use the finest granularity supported by available evidence. A coarse fact still answers a fine-grained question conservatively.

\textbf{Alias resolution is part of identity, not a convenience.} Real systems place indirection between a consumer-facing name and its definition. Examples include re-export layers, registry modules, dependency-injection bindings, and \code{export ... as ...}. The normalization function must follow these chains to the \emph{leaf definition}. Otherwise, one artifact acquires multiple identities, subsumption breaks, and a leaf change becomes invisible to a claim anchored on an alias. The function therefore assigns identity at the leaf and treats every alias as another name for that identity. In the testbed, such indirection was the most frequently reported source of difficulty (Section~5). It therefore belongs in the model rather than an implementation note.

\begin{figure}[t]

\centering

\includegraphics[width=0.98\textwidth]{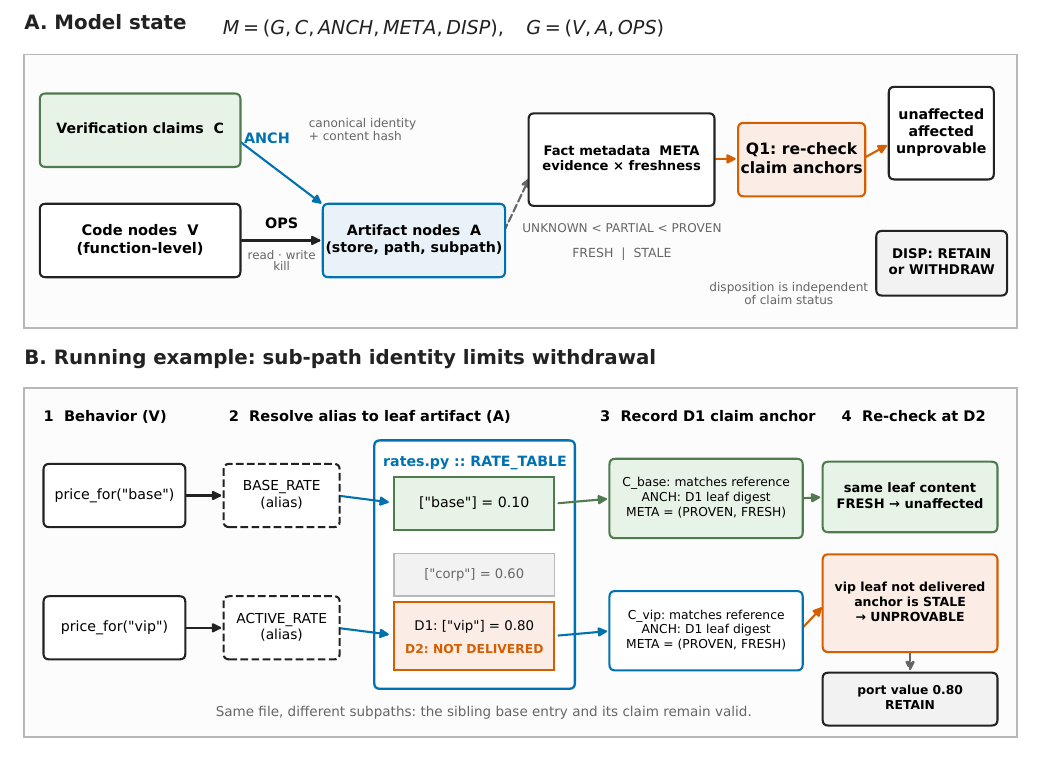}

\caption{Overview of the EA-Graph model and an illustrative claim re-check.}

\label{fig:figure2_eagraph_framework}

\end{figure}

\textbf{Running example.} Figure~\ref{fig:figure2_eagraph_framework} instantiates the complete state without exposing an experimental world. Read the lower panel from left to right:

\begin{enumerate}

\item \textbf{Behavior.} \code{price\_for("base")} and \code{price\_for("vip")} are separate code nodes. Both happen to read values stored in \code{rates.py}.

\item \textbf{Identity resolution.} The consumer names \code{BASE\_RATE} and \code{ACTIVE\_RATE} are aliases, not artifact identities. Normalization follows each alias to a different leaf: \code{RATE\_TABLE["base"]} and \code{RATE\_TABLE["vip"]}. Sharing one table and file therefore does not make them the same artifact.

\item \textbf{D1 anchoring.} Verification at D1 creates two claims. \code{C\_base} records a digest of value \code{0.10}; \code{C\_vip} records a digest of value \code{0.80}. Both enter with META \code{(PROVEN, FRESH)} because their evidence is deterministic and the anchored content still matches.

\item \textbf{D2 re-check.} D2 delivers the \code{base} leaf unchanged, so \code{C\_base} remains FRESH and the withdrawal query returns \textbf{unaffected}. The new \code{vip} leaf is not delivered. Its D1 anchor is therefore STALE, and the claim outcome is \textbf{UNPROVABLE} rather than affected: no replacement exists to check. DISP remains \textbf{RETAIN}, so the port keeps the last verified value \code{0.80} rather than guessing a replacement or erasing working code.

\end{enumerate}

The last step separates three decisions. STALE is a freshness observation, UNPROVABLE is a claim outcome, and RETAIN is an artifact disposition. File-level invalidation would withdraw both claims. Leaf identity withdraws only the claim whose entry changed.

\textbf{OPS records effect, not value flow.} An edge (v, a, read) states that code node \emph{v} consumes artifact \emph{a}. The write relation states that \emph{v} defines \emph{a}, whereas kill denotes deletion, renaming, or a schema-breaking change. This layer differs from statement-level data-flow analysis. Def-use chains, slicing, and taint propagation answer questions \emph{within} a procedure's value graph. Agent-facing examples include ARISE's intra-procedural slicing \citep{arise2026} and RepoAudit's path-wise data-flow facts \citep{repoaudit2025}. OPS instead answers questions across persistent system state: which code reads a key, rewrites a registry entry, or deletes a column. The layers can compose. Value-flow analysis can refine an effect edge to statement precision, while an effect layer can select procedures for slicing. The contribution here is the effect layer, its identity discipline, and its evidence semantics. It is not a reinvention of def-use analysis.

The same distinction applies to symbol-graph representations for agents \citep{codexgraph2025,repograph2025,codeplan2024}. They capture definitions and references among \emph{code symbols}. A constant table is a symbol, but the entry read by a behavior is not. Drift can occur at that entry.

Content-addressed build systems also invalidate content after a hash change. However, they typically invalidate whole files to trigger a \textbf{rebuild}. EA-Graph instead invalidates artifact sub-paths to \textbf{withdraw a claim}. It also operates without a build graph or upstream version history, as described in Section~4.

\subsection{Evidence and freshness}\label{sec:3.2}

Every fact in the graph carries a pair \textbf{(evidence, freshness)}. The two are independent lattices, and keeping them apart is what lets the model say the two different things an agent needs to hear.

\textbf{Evidence} is drawn from \textbf{UNKNOWN < PARTIAL < PROVEN} and is a statement about \emph{grounds}:

\begin{itemize}

\item \textbf{PROVEN} --- deterministic extraction (syntactic facts, hash-verified captures) or verified observation.

\item \textbf{PARTIAL} --- proposed on incomplete grounds. Every LLM-proposed edge enters here, as does every deterministic fact whose scope had to be widened: a computed key yields a PARTIAL edge on the \emph{whole} container, never a guessed key.

\item \textbf{UNKNOWN} --- a boundary crossing with no linkage evidence: the writer and the reader of a cross-process artifact are both known, and their identity linkage is not.

\end{itemize}

\textbf{Freshness} is \textbf{FRESH} or \textbf{STALE} and describes \emph{time}. A fact becomes STALE when its source has changed since extraction. A content hash over the anchored span, rather than the whole file, detects this change.

Three rules give these states force.

\begin{enumerate}

\item \textbf{Query contract.} A derived answer carries the \textbf{meet} of the evidence along its derivation path. One weak premise grades the whole conclusion. The meet also identifies the next premise to strengthen.

\item \textbf{Refusal.} If any fact on the path is STALE, the query is \textbf{refused}, with a rebuild obligation naming the drifted anchors --- at whatever evidence grade it would otherwise have carried. Freshness is not a caveat to be attached to an answer the agent is going to use anyway.

\item \textbf{LLM quarantine.} No model output enters at PROVEN. Promotion requires deterministic re-extraction or observation.

\end{enumerate}

Because the lattices are independent, their product forms a lattice with a componentwise meet. Their separation therefore does not complicate rule 1.

\subsection{Unprovability and anchor completeness}\label{sec:3.3}

\textbf{UNPROVABLE is a terminal state, not a low grade.} A claim is UNPROVABLE when a required artifact has changed and its new content is unavailable. The upstream may have withheld the content, placed it beyond an accessible boundary, or omitted it from delivery. No further work within the agent's reach can settle the claim. Modeling this state as low confidence invites unsupported action. A terminal state instead requires the agent to report the limitation.

\textbf{Claim status and artifact disposition are separate.} An artifact with an UNPROVABLE claim is not necessarily wrong; it has become \emph{unverifiable}. The store therefore records claim status independently from artifact disposition. The last verified content may be retained with its prior verification record and the reason that its reference is unavailable. Alternatively, it may be withdrawn. Collapsing these decisions encourages an agent to destroy working code merely to express uncertainty. Sections 5.4 and 6.3 show how the distinction affected observed behavior.

\textbf{Anchor completeness.} A claim about behavior \emph{v} is anchored to a set anchors(v) \ensuremath{\subseteq} A. Detection of drift is only as good as that set:

\begin{quote}

anchors(v) is \textbf{complete} iff \{ a : (v, a, read) \ensuremath{\in} OPS \} \ensuremath{\subseteq} \ensuremath{\downarrow}anchors(v),

\end{quote}

where \ensuremath{\downarrow} closes under subsumption. If a read artifact is missing from anchors(v), its drift produces \textbf{no signal}. The claim silently survives a change that invalidates it. This failure is especially serious because it is indistinguishable from correct operation.

The model therefore derives anchors and effect edges through \textbf{two independent paths}. Completeness then becomes checkable rather than assumed. Source extraction produces OPS, while anchoring records what a verification session consulted. The containment relation above can compare the two. Several experimental sessions constructed this cross-check without prompting. One session found a shared helper that no prior session had anchored (Section~5.4).

\subsection{What the model is asked}\label{sec:3.4}

The model exists to serve a small number of queries. They are stated here because they, not the data structure, are what Section~5 measures.

\begin{itemize}

\item \textbf{Q1 --- Withdrawal.} \emph{Which of my recorded claims can I no longer stand behind?} Returns the claims whose anchors are STALE, partitioned into those that can be re-established from what is available and those that are UNPROVABLE. This is the query the evaluation is built around.

\item \textbf{Q2 --- Dependence.} \emph{Which artifacts does this behavior actually use?} The read-set \{ a : (v, a, read) \ensuremath{\in} OPS \}, at sub-path granularity. Q1 is meaningless without it: "affected" can only be decided against the values a behavior uses, not against the files they live in.

\item \textbf{Q3 --- Impact.} \emph{Which behaviors does this change invalidate?} The inverse of Q2 through subsumption, which is what makes the artifact layer worth materializing at all.

\end{itemize}

An honest answer to Q1 has three possible outcomes: \textbf{unaffected}, \textbf{affected}, or \textbf{unprovable}. An affected claim depends on a value or piece of logic that changed. Systems that omit the third outcome force a guess when an upstream withholds required content.

\subsection{Reduced implementation and scope}\label{sec:3.5}

We implemented a \textbf{reduced version of EA-Graph in Python}. The implementation focuses on the withdrawal query (Q1) and its supporting mechanisms; components needed only for other queries are not exercised. It stores one claim per behavior and a set of anchors for each claim. Each anchor combines an artifact identity with a content hash of the span used for verification. Alias chains resolve to their leaf definitions before hashing. Re-checking a claim hashes its anchors again and returns those that changed. A claim with a changed anchor becomes STALE. If the new content is unavailable, the claim becomes UNPROVABLE. Claim status remains separate from artifact disposition, so a claim can be withdrawn without destroying the artifact.

Table~\ref{tab:1} states which model elements the reduced implementation realizes and which the evaluation does \emph{not} touch. Making the scope explicit prevents an evaluation's boundary from disappearing into prose and identifies which parts of the formalism have actually been tested.

\begin{center}

\small

\begin{longtable}{@{}L{0.28\textwidth}L{0.42\textwidth}L{0.22\textwidth}@{}}

\caption{Scope of the reduced EA-Graph implementation.}\label{tab:1} \\

\toprule

\textbf{Model element (\S{})} & \textbf{Realized in the reduced implementation} & \textbf{Evaluated in this paper} \\

\midrule

\endfirsthead

\toprule

\textbf{Model element (\S{})} & \textbf{Realized in the reduced implementation} & \textbf{Evaluated in this paper} \\

\midrule

\endhead

Artifact identity: (store, path, subpath) (Section~3.1) & Yes --- sub-path identity via a single normalization function & \textbf{Yes} \\

Alias resolution to the leaf definition (Section~3.1) & Yes --- re-export chains resolved before identity is assigned & \textbf{Yes} \\

Subsumption order (Section~3.1) & Partial --- container/entry containment only & Indirectly \\

OPS \code{read} edges (Section~3.1) & Yes --- a claim's anchor set is the behavior's read-set & \textbf{Yes} \\

OPS \code{write} / \code{kill} edges (Section~3.1) & Defined, not populated & No --- the task never writes artifacts \\

Freshness, STALE by span hash (Section~3.2) & Yes & \textbf{Yes} \\

Refusal on STALE (Section~3.2) & Yes --- re-check reports drift and withholds the claim & \textbf{Yes} \\

Evidence grade PROVEN (Section~3.2) & Yes & \textbf{Yes} \\

Evidence grades PARTIAL / UNKNOWN (Section~3.2) & No --- extraction is fully deterministic here & \textbf{No} \\

Meet along a derivation path (Section~3.2) & Not needed --- claims are single-hop & \textbf{No} \\

UNPROVABLE as a terminal state (Section~3.3) & Yes & \textbf{Yes} \\

Claim status separate from artifact disposition (Section~3.3) & Yes & \textbf{Yes} (Section~5.4, Section~6.3) \\

Anchor completeness as a checkable property (Section~3.3) & Anchors and effect edges are derived independently & Observationally (Section~5.4) \\

\bottomrule

\end{longtable}

\end{center}

As Table~\ref{tab:1} shows, the evaluated path covers sub-path identity, alias normalization, freshness, refusal, and unprovability. It does not cover general evidence grading, write/kill effects, or multi-hop meets. Two other components are also outside the study. A \textbf{plan layer} could record intended changes by reference to G, but this layer is neither defined nor evaluated here. Nothing in Section~5 depends on it. The model also permits \textbf{incremental maintenance} under change, which may matter at repository scale. The reduced implementation hashes all anchors eagerly, so the evaluation supports no efficiency claim. Section 7 presents both components as future work rather than contributions.

\section{A Generated Testbed and Why an Existing Codebase Is Insufficient}\label{sec:4}

The claim under test concerns \emph{judgment}: after an upstream changes, which previously verified claims can the agent no longer support? Testing this claim requires known answers that cannot be recovered without reasoning about the code. The environment must also vary independently across replications. No existing open-source project satisfies these requirements. Section 4.1 identifies five obstacles, and \S{}Section~4.2--4.5 describe the corresponding controls.

\subsection{Five obstacles and their design consequences}\label{sec:4.1}

Table~\ref{tab:2} maps each validity threat in an existing codebase to a testbed control. These controls are interdependent. Known ground truth is insufficient if version history reveals the answer. Removing history is also insufficient if a model can recall a public repository.

\begin{center}

\small

\begin{longtable}{@{}L{0.03\textwidth}L{0.19\textwidth}L{0.34\textwidth}L{0.36\textwidth}@{}}

\caption{Obstacles in existing codebases and testbed responses.}\label{tab:2} \\

\toprule

\textbf{\#} & \textbf{Obstacle in a real project} & \textbf{Why it is fatal} & \textbf{Design response} \\

\midrule

\endfirsthead

\toprule

\textbf{\#} & \textbf{Obstacle in a real project} & \textbf{Why it is fatal} & \textbf{Design response} \\

\midrule

\endhead

1 & \textbf{No ground truth} & For a given upstream change, no real project can enumerate \emph{which previously verified behavioral claims are now unprovable}. That requires a complete behavior \ensuremath{\leftrightarrow} artifact \ensuremath{\leftrightarrow} value map, which no repository maintains. Without it the primary measure cannot be computed at all. & Generate the entire world from an intermediate representation (Section~4.2), which yields exact provenance and a mechanically derived acceptance suite. \\

2 & \textbf{Version control is itself a perfect change oracle} & Where history exists, one diff answers the research question directly and dominates every memory design. This is not hypothetical: in an early experimental round, \textbf{19 of 24 sessions} used the previous commit as a value-level change oracle, and every condition converged. & Deliver upstream as an \textbf{unversioned drop} (Section~4.4): files are replaced in place, the reference tree is excluded from version control, and timestamps are normalized. \\

3 & \textbf{Training-data contamination} & Popular repositories are in the model's weights. A correct answer may be recalled rather than derived, and any benchmark built on them decays as it is scraped. & Every identifier is freshly synthesized for that world, so nothing can be recalled (Section~4.2); the testbed is released under access control with a canary (Section~4.5). \\

4 & \textbf{Confounded difficulty, no replication} & A repository gives one instance. Difficulty cannot be held constant, so an effect and a hard example cannot be told apart. & Seeded generation produces worlds with a matched drift budget (Section~4.3), so replications differ in content but not in difficulty. \\

5 & \textbf{No way to withhold} & To test whether an agent will \emph{refuse} rather than guess, the upstream must sometimes fail to ship the new content. Real upstreams do not redact on request. & A partial-delivery pass replaces selected definitions with an explicit \emph{withheld} marker and records the resulting unprovable set as ground truth (Section~4.3). \\

\bottomrule

\end{longtable}

\end{center}

Obstacle 2 in Table~\ref{tab:2} deserves emphasis because it may appear artificial. An agent that can diff the previous reference revision does not need a memory of what it verified. The diff provides a more direct record. However, that route is unavailable for vendor SDK drops, generated code, exported asset pipelines, and other upstreams delivered as directories rather than repositories. Removing history reproduces the condition under which the problem arises; it does not weaken the baseline.

\subsection{The generated world}\label{sec:4.2}

A world is generated from a seed. A specification pass emits an intermediate representation of \textbf{96 behaviors} across \textbf{12 modules}. Each behavior is a short pipeline drawn from \textbf{16 operation types}. These types include table lookups, banded steps, affine and modular arithmetic, windowed sums, parity gates, alias substitutions, format padding, and fault conditions. Composition ensures that each result depends on a small, exactly known set of data definitions.

From that one representation the generator emits three artifacts that agree by construction:

\begin{itemize}

\item a \textbf{Python reference} --- the documentation the agent reads, split into a core layer of behavior implementations and a periphery layer of data definitions;

\item a \textbf{TypeScript port}, produced by an oracle renderer, which is the artifact the agent maintains;

\item an \textbf{acceptance suite} of \textbf{576 vectors} (six per behavior) with expected outputs computed from the representation.

\end{itemize}

Three properties of this construction matter for the argument.

\textbf{The reference cannot be executed.} It imports a compiled extension that is not present. This constraint removes differential testing as a route to the answer. The agent cannot run the reference against the port and compare their outputs; it must reason about source code. Without this constraint, the task would reduce to running a test harness.

\textbf{Consumers do not name definitions directly.} A registry layer re-exports each data definition under a different name. Thus, a behavior never uses the identifier under which a value is defined. This design reproduces the indirection that identity normalization must resolve (Section~3.1). A naive name-matching approach will silently miss such drift.

\textbf{Nothing in the world is recallable.} Module names, behavior names, and identifiers are generated separately for each world as fresh synthetic names. A model that has never seen the world cannot have memorized any part of the answer.

The acceptance suite is never visible to the agent. It remains outside the workspace, runs only after the session, and provides no feedback (Section~4.4).

\subsection{Drift and partial delivery}\label{sec:4.3}

Each world has three generated states: a \textbf{base}, a first drop \textbf{D1}, and a second drop \textbf{D2}. The port is constructed and verified against D1, where the agent's claims are true. D2 is the changed drop presented during evaluation. Every replication uses the same drift budget:

\begin{itemize}

\item \textbf{Data drift.} Roughly eighteen data definitions change value between D1 and D2 --- table entries added or altered, a scan order reversed, a mode flag flipped, a padding width changed. These are the changes that leave the code compiling and the file list unchanged.

\item \textbf{Logic drift.} One behavior has two adjacent operations transposed in the reference. For example, a parity gate may run before rather than after a fault check. Transposition is restricted to operations of the same type class, ensuring a meaningful swap. The generator accepts the swap only if it changes at least one output. Logic drift is included because data-value comparison alone cannot detect it.

\item \textbf{Partial delivery.} Three definitions are replaced in D2 by an explicit \emph{withheld} marker stating that the content was not shipped. By construction, every behavior that reads a withheld definition is unprovable. Its correctness cannot be established from the delivered material. The generator records the resulting set as ground truth.

\end{itemize}

Two build-time gates run before a world is used. An \textbf{oracle gate} confirms a score of 576/576 for the rendered port against D1 and the expected lower score against D2. Every lost point is therefore attributable to drift rather than a generator defect. A \textbf{mechanism gate} tests regeneration from the \emph{delivered} D2 reference. Across worlds, entry scores are 516--526 out of 576, while the regeneration ceiling is 561--570. The residual gap corresponds exactly to withheld material. Mechanical regeneration therefore cannot complete the task; some claims require an explicit judgment of unprovability.

\subsection{Session protocol and revision history}\label{sec:4.4}

An episode is a \textbf{single session with no feedback}. The prompt states that an updated reference drop has arrived, the reference is read-only and non-executable, and earlier revisions are unavailable. The agent must bring the port to a state it can support. The prompt also discloses a hidden acceptance suite of 576 checks, but no score is reported during the session. At the end, the agent files one status line per behavior: \textbf{unaffected}, \textbf{affected}, or \textbf{unprovable}.

\begin{figure}[t]

\centering

\includegraphics[width=0.98\textwidth]{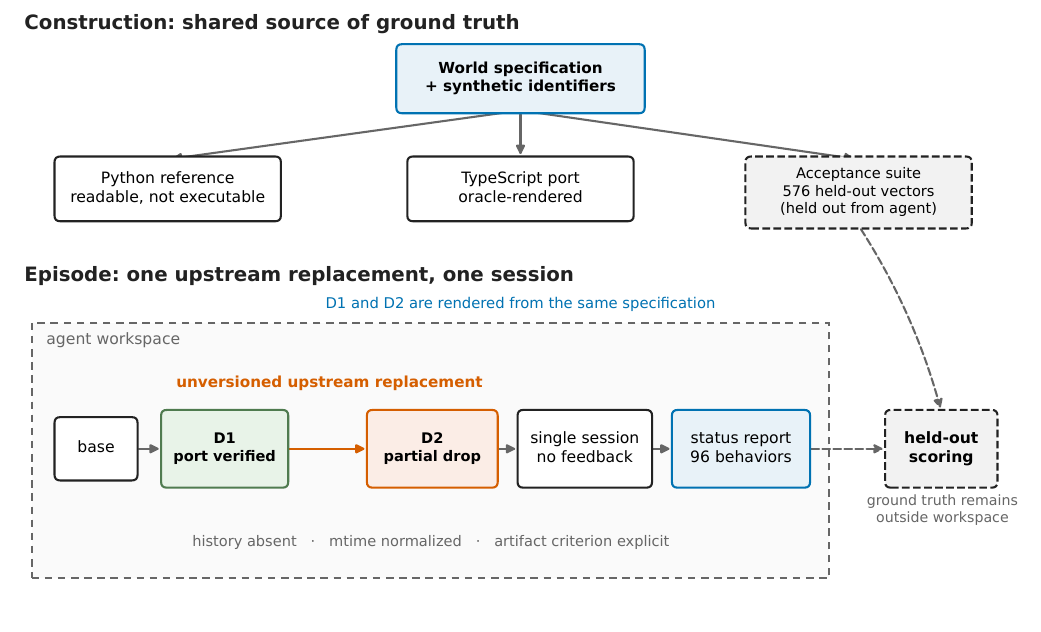}

\caption{Generated-testbed construction and single-session protocol.}

\label{fig:figure3_testbed_protocol}

\end{figure}

Figure~\ref{fig:figure3_testbed_protocol} separates testbed construction from the episode visible to the agent. The upper path derives the reference, D1-verified port, ground truth, and held-out acceptance suite from one specification. The lower path exposes only the port and an unversioned D2 reference during a no-feedback session. The scorer and earlier reference state remain outside the workspace. This separation prevents leakage through evaluation feedback or a direct revision diff.

The protocol reached this form through several revisions. Each revision closed a route to the answer that bypassed the intended reasoning. These closures are methodologically important because such shortcuts can invalidate an evaluation of agent memory. Table~\ref{tab:3} lists three routes observed in earlier revisions and the corresponding controls.

\begin{center}

\small

\begin{longtable}{@{}L{0.20\textwidth}L{0.34\textwidth}L{0.38\textwidth}@{}}

\caption{Shortcut routes and protocol closures.}\label{tab:3} \\

\toprule

\textbf{Route discovered} & \textbf{What it gave away} & \textbf{Closure} \\

\midrule

\endfirsthead

\toprule

\textbf{Route discovered} & \textbf{What it gave away} & \textbf{Closure} \\

\midrule

\endhead

\textbf{Version history} & The previous revision of the reference, hence a perfect value-level diff. Used by 19 of 24 sessions in an early round. & The reference tree is excluded from version control and replaced in place; only the port is versioned. \\

\textbf{File modification times} & A drop rewrites only the files it changes, so a directory listing is a free file-level changelog. One session used exactly this. & All reference timestamps are normalized to a single value --- and to the \emph{oldest} in the tree, not merely a uniform one: a uniform but recent stamp is itself a signal, and one session read it as "everything was rewritten." \\

\textbf{Ambiguity in the question} & Asking which behaviors were affected by "something they depend on" is \emph{true at file granularity} whenever any symbol in an imported file changed. Sessions answered that reading almost perfectly while being scored against artifact-granularity truth. & The task states the criterion explicitly, in wording identical across conditions: judge each behavior by the specific values and logic it actually uses, not by the files those live in. \\

\bottomrule

\end{longtable}

\end{center}

The third closure in Table~\ref{tab:3} yielded a broader lesson. Before the clarification, the measured gap between conditions was near zero. Afterward, with the same memory, tools, and model, the anchored condition's median rose from 0.67 to 1.000. \textbf{An artifact-granularity tool is ineffective when the question is posed at file granularity.} The task definition must encode the methodology rather than leave it only in the instrument. Section 5 reports both granularities for every session, making the distinction auditable.

One discovered route was \textbf{not} closed: stale compiled bytecode left in the reference tree encoded an \emph{older} revision. Section 6.3 analyzes it because measuring its strength turned it into a finding rather than merely a defect.

\subsection{Controlled release}\label{sec:4.5}

The value of this testbed depends on its answers not being in anyone's training data. Publishing the generated worlds and their ground truth would destroy that property for every subsequent user, including for the crawlers that would ingest it within weeks of posting.

The release is therefore split across two public repositories. The testbed structure contains the \textbf{generator design, drift and withholding procedures, session protocol, measure definitions, schemas, and aggregate analysis code} \citep{eagraphbenchmark2026}. The companion research record contains the disclosure-safe study history, preregistration decisions, results, incident register, and claim audit \citep{eagraphrecord2026}. Together, these materials support method assessment, design reproduction, and construction of new instances. The generated worlds, acceptance suites, and ground-truth files are available \textbf{on justified request} from the corresponding author. Each approved archive includes a requester-specific canary string to detect later public or training-data leakage.

This policy balances reproducibility against benchmark integrity. All material needed to assess the design and reasoning is public. Content that would expose hidden answers remains controlled so that later evaluations remain meaningful.

\section{Evaluation}\label{sec:5}

\subsection{Design, conditions, and measures}\label{sec:5.1}

\textbf{Conditions.} Each session uses one of the three conditions in Table~\ref{tab:4}. The conditions differ only in the information carried across sessions. Task text, tools, world, and reference drop remain identical. This design isolates content-anchored memory from prose handoff and reconstruction without memory.

\begin{center}

\small

\begin{longtable}{@{}L{0.17\textwidth}L{0.75\textwidth}@{}}

\caption{Experimental memory conditions.}\label{tab:4} \\

\toprule

\textbf{Condition} & \textbf{What the workspace carries} \\

\midrule

\endfirsthead

\toprule

\textbf{Condition} & \textbf{What the workspace carries} \\

\midrule

\endhead

\textbf{ANCHOR} & The reduced EA-Graph store of Section~3.5: one claim per behavior, anchored to the artifact content it was verified against, with staleness detected by re-hashing. \\

\textbf{PROSE} & A notes file with one line per verified behavior, recording in prose that the behavior was checked and where. It represents a lightweight handoff record without content anchors. \\

\textbf{NONE} & No cross-session record at all. A floor, not a straw man: it measures what the model can reconstruct from the workspace alone. \\

\bottomrule

\end{longtable}

\end{center}

The crucial contrast in Table~\ref{tab:4} is not record coverage. Before each session, ANCHOR and PROSE receive complete and correct records for all 96 behaviors. Both records originate from the same verification event. The conditions differ in record \emph{content}, not initial coverage.

\textbf{Rounds.} Both rounds use the same seven clean worlds. Each world is evaluated under three conditions with \code{claude-haiku-4-5-20251001} (21 sessions) and then with \code{claude-sonnet-5} (21 sessions). Both rounds were dispatched on 2026-08-01. The analysis therefore covers \textbf{42 sessions}, seven unique worlds, and 14 model-world instances. One additional Haiku world was excluded in full because its generated base template contained files from an earlier process (Section~6.4). None of its three sessions appears in the analysis.

Both rounds were preregistered before dispatch. The Haiku registration fixed the primary measure, effect direction, continuation rule, and stopping criteria. However, the exact Wilcoxon test was selected after the runs, so its \emph{p}-values are not fully confirmatory. The Sonnet registration fixed both within-round contrasts and the exact test. It designated cross-model convergence thresholds as descriptive analyses only.

The tiers also differ in vendor-documented inference affordances. Anthropic describes Haiku 4.5 as its fastest near-frontier model and Sonnet 5 as a higher-intelligence model for coding and agents \citep{anthropicmodels2026}. The API documentation lists a 200k-token context window and 64k maximum output for \code{claude-haiku-4-5-20251001}. For \code{claude-sonnet-5}, it lists a default 1M-token context window, 128k maximum output, and adaptive thinking by default \citep{anthropicmigration2026}. These specifications motivate treating Sonnet as the stronger, larger-context comparison. However, model generation, training, tokenization, and inference policy also differ. The experiment cannot identify context-window size as a causal mechanism.

\textbf{Primary measure.} Each session classifies all 96 behaviors. The primary measure is \textbf{F1 over the not-OK set}, defined as the union of \emph{affected} and \emph{unprovable} behaviors. It captures both the recall and precision of claim withdrawal. A missing report is scored as classifying every behavior as unaffected rather than treated as missing data.

\textbf{File-granularity diagnostic.} The same classifications are also scored against file-level ground truth. Under this definition, a behavior is affected whenever any symbol changes in a file from which it imports. This score is a diagnostic, not a fallback measure. Because the task explicitly requires artifact granularity (Section~4.4), a high file-level score indicates that a session answered a different question. Reporting both scores makes this error directly observable. The file-level not-OK set contains 81--91 of 96 behaviors, whereas the artifact-level set contains 14--18.

\textbf{Secondary measures.} Secondary measures include the terminal acceptance score out of 576 and its decomposition (Section~5.4). They also include \emph{fabrication}, defined as substantive content written for a withheld definition, and \emph{emptying}, defined as replacement of such a definition with an empty value. For ANCHOR sessions, an external trace records whether the store was consulted and whether its reported set matched the filed classifications. This trace is not visible within the session.

\textbf{Statistics.} Within each round, conditions are paired by world. Comparisons use an exact two-sided Wilcoxon signed-rank test computed by enumeration, with ties removed. With seven or eight non-tied pairs, the minimum attainable two-sided \emph{p} is 0.0156 or 0.0078, respectively. Cross-round comparisons align the shared worlds descriptively. They are not a confirmatory factorial analysis of model by condition.

\subsection{Main result}\label{sec:5.2}

\textbf{Haiku round.} Table~\ref{tab:5} reports per-world F1 over the not-OK set. Table~\ref{tab:6} gives the paired signed-rank comparisons. Together, they separate consistency from magnitude. ANCHOR wins in every world despite one low absolute score, whereas the two controls exchange wins and remain low.

Tables~\ref{tab:5} and~\ref{tab:6} show that ANCHOR outscored both controls in all seven clean worlds. The PROSE-versus-NONE contrast shows no detectable difference (\emph{p} = 1.000). This null result does not establish that prose and no memory are equivalent. The primary metric and direction were preregistered, while the exact test form was selected post hoc (Section~5.1).

\textbf{Sonnet round.} Table~\ref{tab:7} reports the corresponding per-world results, and Table~\ref{tab:8} gives the paired comparisons. ANCHOR remains perfect, but each control also reaches 1.000 in four worlds. This ceiling sharply reduces the number of non-tied pairs.

\begin{center}

\small

\begin{longtable}{@{}lrrr@{}}

\caption{Haiku per-world classification F1.}\label{tab:5} \\

\toprule

\textbf{World} & \textbf{ANCHOR} & \textbf{PROSE} & \textbf{NONE} \\

\midrule

\endfirsthead

\toprule

\textbf{World} & \textbf{ANCHOR} & \textbf{PROSE} & \textbf{NONE} \\

\midrule

\endhead

1 & \textbf{1.000} & 0.300 & 0.300 \\

2 & 0.364 & 0.000 & 0.286 \\

3 & \textbf{1.000} & 0.000 & 0.000 \\

4 & \textbf{1.000} & 0.815 & 0.286 \\

5 & \textbf{1.000} & 0.303 & 0.133 \\

6 & \textbf{1.000} & 0.270 & 0.000 \\

7 & \textbf{1.000} & 0.000 & 0.410 \\

\textbf{median} & \textbf{1.000} & 0.270 & 0.286 \\

\bottomrule

\end{longtable}

\end{center}

\begin{center}

\small

\begin{longtable}{@{}L{0.40\textwidth}rrrrr@{}}

\caption{Haiku paired condition comparisons.}\label{tab:6} \\

\toprule

\textbf{Comparison} & \textbf{Wins} & \textbf{Losses} & \textbf{Ties} & \textbf{Effective \emph{n}} & \textbf{Exact \emph{p}} \\

\midrule

\endfirsthead

\toprule

\textbf{Comparison} & \textbf{Wins} & \textbf{Losses} & \textbf{Ties} & \textbf{Effective \emph{n}} & \textbf{Exact \emph{p}} \\

\midrule

\endhead

ANCHOR vs PROSE & \textbf{7} & 0 & 0 & 7 & \textbf{0.0156} \\

ANCHOR vs NONE & \textbf{7} & 0 & 0 & 7 & \textbf{0.0156} \\

PROSE vs NONE & 3 & 2 & 2 & 5 & 1.000 \\

\bottomrule

\end{longtable}

\end{center}

\begin{center}

\small

\begin{longtable}{@{}lrrr@{}}

\caption{Sonnet per-world classification F1.}\label{tab:7} \\

\toprule

\textbf{World} & \textbf{ANCHOR} & \textbf{PROSE} & \textbf{NONE} \\

\midrule

\endfirsthead

\toprule

\textbf{World} & \textbf{ANCHOR} & \textbf{PROSE} & \textbf{NONE} \\

\midrule

\endhead

1 & \textbf{1.000} & 0.970 & 0.694 \\

2 & \textbf{1.000} & 0.971 & \textbf{1.000} \\

3 & \textbf{1.000} & \textbf{1.000} & 0.970 \\

4 & \textbf{1.000} & \textbf{1.000} & \textbf{1.000} \\

5 & \textbf{1.000} & 0.718 & \textbf{1.000} \\

6 & \textbf{1.000} & \textbf{1.000} & \textbf{1.000} \\

7 & \textbf{1.000} & 0.970 & 0.970 \\

\textbf{median} & \textbf{1.000} & 0.971 & \textbf{1.000} \\

\bottomrule

\end{longtable}

\end{center}

\begin{center}

\small

\begin{longtable}{@{}L{0.40\textwidth}rrrrr@{}}

\caption{Sonnet paired condition comparisons.}\label{tab:8} \\

\toprule

\textbf{Comparison} & \textbf{Wins} & \textbf{Ties} & \textbf{Effective \emph{n}} & \textbf{Exact \emph{p}} & \textbf{Floor at this \emph{n}} \\

\midrule

\endfirsthead

\toprule

\textbf{Comparison} & \textbf{Wins} & \textbf{Ties} & \textbf{Effective \emph{n}} & \textbf{Exact \emph{p}} & \textbf{Floor at this \emph{n}} \\

\midrule

\endhead

ANCHOR vs PROSE & 4 & 3 & 4 & 0.125 & 0.125 \\

ANCHOR vs NONE & 3 & 4 & 3 & 0.250 & 0.250 \\

PROSE vs NONE & 2W/2L & 3 & 4 & 1.000 & 0.125 \\

\bottomrule

\end{longtable}

\end{center}

Tables~\ref{tab:7} and~\ref{tab:8} show that ANCHOR scores 1.000 in all seven worlds, with no variance. The Wilcoxon signed-rank test excludes zero-difference pairs, so not all seven worlds contribute to every contrast. Three ANCHOR--PROSE ties leave an effective \emph{n} of 4. Four ANCHOR--NONE ties leave \emph{n} = 3, and three PROSE--NONE ties leave \emph{n} = 4. At these sample sizes, the minimum attainable two-sided exact \emph{p}-values for the preregistered ANCHOR contrasts are 0.125 and 0.250. Every non-tied difference favors ANCHOR, but too few such differences remain for the exact test to reach 0.05. The preregistered Sonnet hypotheses are therefore \textbf{not supported}. This outcome is not evidence of no effect. Under the observed ceiling, the design cannot distinguish no effect from an effect that yields too few non-tied worlds to resolve.

Across all 14 clean model-world instances, ANCHOR is descriptively never below PROSE and exceeds it by at least 0.25 in 7 instances. This pooled statement is not an inferential test.

\begin{figure}[t]

\centering

\includegraphics[width=0.98\textwidth]{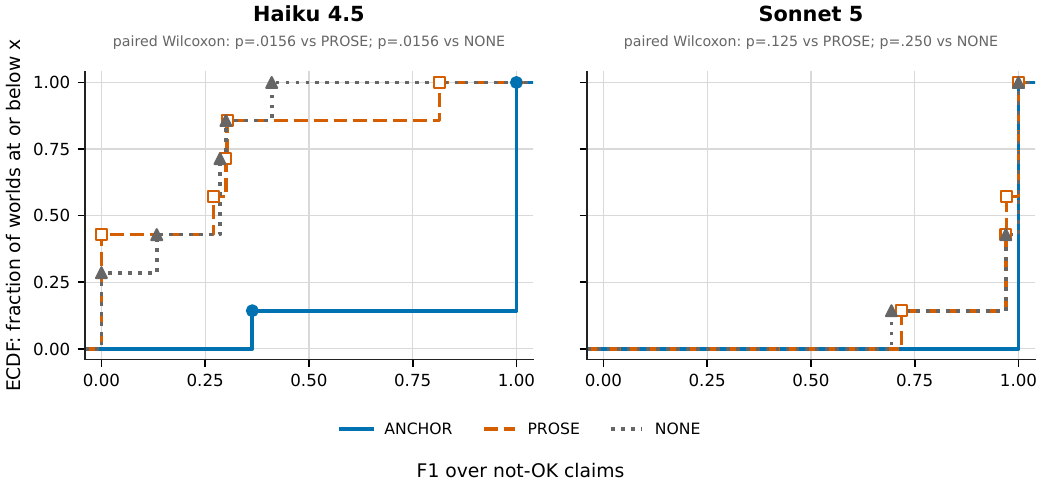}

\caption{ECDFs of classification F1 by model and memory condition.}

\label{fig:figure4_main_results}

\end{figure}

Figure~\ref{fig:figure4_main_results} displays the per-world values as empirical cumulative distributions. In the Haiku panel, the ANCHOR curve lies to the right of both controls across the distribution. This view shows the consistent directional advantage rather than only the median. In the Sonnet panel, all three curves cluster near 1.0. Their overlapping steps reveal the ceiling and ties that reduce the effective sample sizes in Table~\ref{tab:8}. The axes extend beyond 0 and 1 only to show boundary masses clearly.

\subsection{Exploratory capability-gap pattern}\label{sec:5.3}

The Sonnet preregistration specified two descriptive convergence criteria. Both were met. The median ANCHOR-minus-PROSE difference fell from +0.700 in Haiku to +0.029 in Sonnet, while Sonnet with NONE reached median F1 = 1.000. For the three worlds with identical timestamp procedures across rounds (1, 6, and 7), the corresponding values were +0.730, +0.030, and 0.970.

Table~\ref{tab:9} summarizes each condition--model distribution, while Table~\ref{tab:10} preserves the seven-world pairing for the comparisons used to motivate the exploratory capability-gap hypothesis. The summary should not be read as an equivalence test: equal medians can coexist with different world-level outcomes.

The bold rows in Table~\ref{tab:10} motivate the capability-gap hypothesis. For ANCHOR\ensuremath{\cdot}Haiku versus NONE\ensuremath{\cdot}Sonnet, the median difference is zero, with three wins, one loss, and three ties (\emph{p} = 0.75). This result does not establish equivalence. The comparison contains only seven pairs and lacks a prespecified equivalence margin. It shows only that no difference was detected on this bounded judgment task, despite Haiku's lighter vendor positioning and smaller documented context window. Within Haiku, ANCHOR is associated with a large improvement over PROSE. Without anchors, PROSE\ensuremath{\cdot}Haiku falls below NONE\ensuremath{\cdot}Sonnet in all seven shared worlds. ANCHOR\ensuremath{\cdot}Haiku instead has the same median as the Sonnet conditions and overlapping world-level scores. Session artifacts suggest a possible mechanism (Section~6.1). Sonnet controls repeatedly reconstructed an artifact baseline within the session, whereas Haiku controls rarely completed that procedure.

\begin{center}

\small

\begin{longtable}{@{}L{0.43\textwidth}rrr@{}}

\caption{Cross-model descriptive classification summary.}\label{tab:9} \\

\toprule

\textbf{Condition \ensuremath{\times} model} & \textbf{median} & \textbf{mean} & \textbf{min} \\

\midrule

\endfirsthead

\toprule

\textbf{Condition \ensuremath{\times} model} & \textbf{median} & \textbf{mean} & \textbf{min} \\

\midrule

\endhead

\textbf{ANCHOR, Haiku} & \textbf{1.000} & 0.909 & 0.364 \\

PROSE, Haiku & 0.270 & 0.241 & 0.000 \\

NONE, Haiku & 0.286 & 0.202 & 0.000 \\

ANCHOR, Sonnet & 1.000 & 1.000 & 1.000 \\

PROSE, Sonnet & 0.971 & 0.947 & 0.718 \\

NONE, Sonnet & 1.000 & 0.948 & 0.694 \\

\bottomrule

\end{longtable}

\end{center}

\newpage

\begin{center}

\small

\begin{longtable}{@{}L{0.22\textwidth}L{0.24\textwidth}L{0.22\textwidth}L{0.22\textwidth}@{}}

\caption{Exploratory matched-world comparisons.}\label{tab:10} \\

\toprule

\textbf{Comparison (7 matched worlds)} & \textbf{W / L / T} & \textbf{median diff} & \textbf{\emph{p}} \\

\midrule

\endfirsthead

\toprule

\textbf{Comparison (7 matched worlds)} & \textbf{W / L / T} & \textbf{median diff} & \textbf{\emph{p}} \\

\midrule

\endhead

\textbf{ANCHOR\ensuremath{\cdot}Haiku vs NONE\ensuremath{\cdot}Sonnet} & \textbf{3 / 1 / 3} & \textbf{0.000} & \textbf{0.75} \\

ANCHOR\ensuremath{\cdot}Haiku vs ANCHOR\ensuremath{\cdot}Sonnet & 0 / 1 / 6 & 0.000 & 1.00 \\

ANCHOR\ensuremath{\cdot}Haiku vs PROSE\ensuremath{\cdot}Sonnet & 3 / 1 / 3 & 0.000 & 0.75 \\

\textbf{PROSE\ensuremath{\cdot}Haiku vs NONE\ensuremath{\cdot}Sonnet} & \textbf{0 / 7 / 0} & -0.730 & \textbf{0.0156} \\

\textbf{ANCHOR\ensuremath{\cdot}Haiku vs PROSE\ensuremath{\cdot}Haiku} & \textbf{7 / 0 / 0} & +0.700 & \textbf{0.0156} \\

\bottomrule

\end{longtable}

\end{center}

This pattern suggests that structured verification-claim memory \textbf{may narrow the observed capability gap} by externalizing re-derivation. It does not show that Haiku and Sonnet are equivalent, that memory replaces model capability, or that context-window size caused the pattern. The comparisons span separate rounds and agent executions. Four of seven worlds also differ in timestamp normalization, and no equivalence margin was specified. Thus, the \emph{p}-values in Table~\ref{tab:10} are post-hoc summaries rather than confirmatory cross-model tests.

\subsection{Secondary results}\label{sec:5.4}

\textbf{No session fabricated withheld content.} Across all 42 analyzed sessions, \textbf{fabrication was zero}. Two Haiku sessions replaced a withheld definition with an empty value, thereby propagating the upstream gap into the port. Both sessions also classified the corresponding behaviors as not-OK. No emptying event occurred in the Sonnet round.

\textbf{Classification accuracy is not repair.} Terminal scores are decomposed by acceptance vector relative to the entry state. Outcomes are \emph{fixed} (fail \ensuremath{\rightarrow} pass), \emph{missed} (fail \ensuremath{\rightarrow} fail), \emph{broken}, and \emph{kept}. The broken category needs further separation. When a session explicitly refuses a behavior whose data was withheld, its previously passing vectors begin to fail. This outcome reflects the intended refusal discipline rather than collateral damage. Table~\ref{tab:11} therefore reports repair, damage, and refusal separately.

\begin{center}

\small

\begin{longtable}{@{}L{0.25\textwidth}L{0.20\textwidth}L{0.20\textwidth}L{0.20\textwidth}@{}}

\caption{Repair, damage, and refusal summary.}\label{tab:11} \\

\toprule

\textbf{Condition \ensuremath{\times} model} & \textbf{Median repair rate} & \textbf{Median damage rate} & \textbf{Refusal vectors} \\

\midrule

\endfirsthead

\toprule

\textbf{Condition \ensuremath{\times} model} & \textbf{Median repair rate} & \textbf{Median damage rate} & \textbf{Refusal vectors} \\

\midrule

\endhead

ANCHOR, Haiku & 10.0\% & 0.0\% & 16 \\

PROSE, Haiku & 0.0\% & 0.0\% & 13 \\

NONE, Haiku & 0.0\% & 0.0\% & 15 \\

ANCHOR, Sonnet & 84.6\% & 0.0\% & 7 \\

PROSE, Sonnet & 76.9\% & 0.0\% & 21 \\

NONE, Sonnet & 80.4\% & 0.0\% & 20 \\

\bottomrule

\end{longtable}

\end{center}

Table~\ref{tab:11} shows that the classification contrast does not imply a repair contrast. ANCHOR-Haiku has a 10.0\% median repair rate despite its classification advantage. By comparison, every Sonnet condition repairs most repairable vectors. This decomposition is post hoc. Across all 42 sessions, \textbf{collateral damage occurred exactly once} (Section~6.2); every other apparent loss in the net score was a refusal. ANCHOR\ensuremath{\cdot}Haiku descriptively approaches the Sonnet conditions in classification but not repair. \textbf{Identifying claims to withdraw is distinct from repairing the underlying behavior.} The paper claims only the former.

\textbf{The store was used, and its output usually determined the report.} An external trace records conformance. All \textbf{14/14} ANCHOR sessions consulted the store before editing the port. In \textbf{13/14}, the filed not-OK set was identical to the drifted set reported by the store. In the exception, the store's set matched ground truth exactly, but the session filed a strict subset after reasoning away several results. Its precision was 1.000 and recall was 0.222. \textbf{Following the method does not guarantee acceptance of its output.} This was the only ANCHOR session in either round with a score below 1.000.

\textbf{Granularity is auditable.} ANCHOR sessions score 0.087--0.333 against the file-level truth set, confirming that they answered at the required artifact granularity. The only two sessions above 0.95 at file granularity scored 0.270 and 0.286 against artifact-level truth. They answered a different, coarser question correctly.

\section{Discussion}\label{sec:6}

\subsection{Structured memory and in-session re-derivation}\label{sec:6.1}

The Sonnet controls did not appear to improve by guessing. Their session artifacts repeatedly show the same reconstruction procedure. They parse the reference statically, resolve alias chains to leaf definitions, and compare the resulting values with \textbf{the port's own data file}. Because the port was verified against the preceding drop, it serves as a fingerprint of that earlier state. Several sessions also dispatched read-only auxiliary agents to inspect control flow. One rejected an auxiliary finding after checking it directly.

The same route was available in the Haiku round. Among the 14 clean PROSE and NONE sessions, however, it produced a high classification score only once. The observed distinction is not awareness that comparison is possible. It is the apparent ability to complete repository-wide reconstruction without a result-compromising error.

This behavior supports a mechanism hypothesis for Section~5.3. \textbf{Structured verification-claim memory changes the task from reconstructing an artifact baseline to retrieving and checking a persisted baseline.} This change may help a less capable model approach a stronger model on the bounded judgment task. The evidence does not measure efficiency or causal mediation. It also does not show that memory eliminates the underlying capability difference.

\subsection{Variance and the single catastrophic session}\label{sec:6.2}

ANCHOR's advantage in the Haiku round is consistent in direction but not in magnitude: six worlds score 1.000, and one scores 0.364. The low outlier is the conformance exception in Section~5.4. The store returned the correct set, but the session reasoned away part of it. Repair completion is also bimodal for this condition.

The outlier cannot be attributed specifically to model randomness. \textbf{Each condition was run once per world}, so run-to-run variance and world difficulty are confounded. Distinguishing them requires repeated runs on fixed worlds (Section~7).

One session deserves individual attention. In Haiku world 7, the ANCHOR session correctly classified all 17 drifted behaviors, with precision and recall both equal to 1.000. It then destroyed the port, reducing the score from 520 to 0 out of 576. The session regenerated the entire data file and reordered an operation. Scoring its pre-edit tree confirmed that its edits caused the loss. This case defines the claim boundary sharply. \textbf{A record of what is no longer provable identifies potential damage. It does not prevent an agent from causing additional damage.}

\subsection{What the measures got wrong}\label{sec:6.3}

Three measurement defects in this study produced findings worth reporting because each is a way this kind of evaluation fails quietly.

\textbf{A net repair score penalizes the advocated discipline.} Repair completion, defined as (terminal - entry) / (ceiling - entry), conflates three outcomes: failure to repair, collateral breakage, and correct refusal of unverifiable output. Every session below its entry score had refused rather than broken behavior. After decomposition, collateral damage across the entire study came from one session (Section~6.2). The smaller model's repair rate also fell from the net estimate to 10\%. A measure must treat unprovability as a distinct state when the model does so.

\textbf{A correctly designed tool still requires a correctly framed task.} The first prompt asked which behaviors were affected by "something they depend on." At file granularity, that statement applied to almost every behavior. Sessions answered that interpretation nearly perfectly but were scored against artifact-level truth. An explicit criterion raised ANCHOR's median from 0.67 to 1.000 without changing the store, tools, or model. This change is not a condition-effect result. It shows that \textbf{the task definition must encode the methodology, rather than leave it only in the instrument}.

\textbf{An apparently strong evidence source can be misleading.} Late in the study, a session reconstructed an older reference revision from stale compiled bytecode. Direct measurement quantified the leak. It recovered a state older than the port's verification baseline and therefore flagged changes from an additional drop. Its recall was 1.000 and its precision was 0.55--0.60. Two sessions used the leak and scored 0.694 and 0.718. Their precision values, 0.531 and 0.560, matched the predicted range. Two other sessions found but rejected the bytecode as evidence from an unrelated state; they scored 1.000 and 0.970. The leak is a testbed defect that depresses the control scores. Closing it would therefore strengthen the round-2 null. It also demonstrates that \textbf{evaluating the trustworthiness of evidence is part of the studied discipline}.

\subsection{Threats to validity}\label{sec:6.4}

\textbf{Synthetic world.} Re-derivation is unusually tractable in this testbed. The data layer contains literals recoverable by static parsing, indirection uses pure aliases, the reference fits within one session, and the port completely fingerprints the preceding drop. Reconstruction may be harder in real systems. However, computed identities, cross-process artifacts, and refactoring may also make EA-Graph harder to maintain. These forces operate in opposite directions, and the experiment does not establish their relative effects.

\textbf{One observation per cell.} Each condition was run once per world. World difficulty, sampling variation, and model nondeterminism are therefore confounded. The low Haiku ANCHOR outlier cannot be assigned to one source.

\textbf{Inference status.} The Haiku primary metric and direction were preregistered, but its exact test form was selected after the runs. The Sonnet within-round hypotheses and exact test were preregistered, and both hypotheses failed. The cross-model convergence thresholds were preregistered as descriptive only; the world-aligned comparisons in Section~5.3 remain exploratory.

\textbf{Cross-round comparability and equivalence.} The model tiers ran in separate rounds with different agent executions. Four of seven shared worlds also used different timestamp-normalization procedures. Only worlds 1, 6, and 7 are fully aligned on that detail. No equivalence margin was specified, so non-significant Haiku-versus-Sonnet comparisons do not establish equivalence.

\textbf{Excluded and leaked artifacts.} One additional Haiku world was excluded after files from an earlier process were found in its generated base template. Its three sessions do not appear in the reported analysis. The run and exclusion audit remain in the companion research record. Separately, all retained workspaces contained stale compiled bytecode from an older state. Two control sessions used it and obtained the lower scores predicted by that state. The leak therefore depresses rather than inflates the control scores, but remains a protocol defect. Concurrent Sonnet sessions also shared a scratch directory. Three reported seeing and ignoring other sessions' scripts or diffs. No ground truth was exposed, but session independence was compromised.

\textbf{Cost was not measured.} Transcript persistence failed before the reported rounds completed, leaving token and tool-call totals unavailable. The lookup-versus-reconstruction account in Section~6.1 is a mechanism hypothesis, not an efficiency claim.

\textbf{Model scope and identity.} The study uses one model family and two service tiers: \code{claude-haiku-4-5-20251001} and \code{claude-sonnet-5}. Both were dispatched on 2026-08-01. Unlike the dated Haiku identifier, the Sonnet identifier has no date suffix and may later resolve to a different service snapshot. The ID and dispatch date identify the deployment more precisely than a tier alias alone, but cannot guarantee snapshot-level reproducibility. Results may not generalize to other families, versions, or capability gaps.

\section{Conclusion and Future Work}\label{sec:7}

Records of what an agent has verified are only as durable as their attachment to what was verified. In 42 analyzed sessions over seven clean worlds and 14 model-world instances, ANCHOR outscored both PROSE and NONE in every Haiku world (\emph{p} = 0.0156 for each comparison). PROSE and NONE showed no detectable difference, although this does not establish equivalence. In the Sonnet round, ANCHOR remained perfect, but the preregistered condition contrasts were not significant because the controls frequently reached the ceiling. The supported condition effect is therefore specific to the Haiku round in this testbed.

The cross-model pattern motivates a more ambitious hypothesis, not a second confirmed result. ANCHOR\ensuremath{\cdot}Haiku reached the same median classification F1 as the Sonnet conditions across the seven shared worlds. The Haiku controls remained substantially lower. Together with the observed reconstruction behavior of the Sonnet controls, this pattern suggests that structured verification-claim memory may narrow a capability gap. The memory externalizes an artifact baseline that would otherwise require in-session reconstruction. The study does not establish cross-model equivalence or model substitution.

The boundary is equally important. The result concerns provability judgment, not repair. One session classified every drifted behavior correctly and then destroyed the port. The repair decomposition is post hoc, and cost was not measured. The generated world also makes reconstruction unusually tractable and artifact identity unusually clean.

Future work should repeat runs on fixed worlds and conduct a preregistered factorial study of both model tiers. That study should specify a meaningful gap-narrowing or equivalence criterion. Further work should instrument token and tool costs and evaluate real repositories with difficult identity resolution and incomplete local decisions. Two engineering questions also remain: a plan layer that links intended changes to the graph, and incremental maintenance under repository-scale change.

\bibliographystyle{IEEEtranN}
\bibliography{refs}

\end{document}